\documentclass[preprint,showpacs,preprintnumbers,amsmath,amssymb]{revtex4}

\usepackage{graphicx}
\usepackage{dcolumn}
\usepackage{bm}

\begin{document}

\title{
Evidence for Carrier-Induced High-$T_{\rm C}$ Ferromagnetism\\
 in Mn-doped GaN film}

\author{S. Yoshii$^1$, S. Sonoda$^2$, T. Yamamoto$^1$, T. Kashiwagi$^1$, 
\\ M. Hagiwara$^1$, Y. Yamamoto$^3$, Y. Akasaka$^4$, K. Kindo$^5$, H. Hori$^3$}
\affiliation{
$^1$KYOKUGEN, Osaka University, 1-3 Machikaneyama, Toyonaka, Osaka 560-8531, Japan\\ 
$^2$Department of Electronics and Information Science, Kyoto Institute of Technology, Kyoto 606-8585, Japan   \\ 
$^3$School of Materials Science, Japan Advanced Institute of Science and Technology (JAIST), 1-1 Asahidai Tatsunokuchi, Ishikawa 923-1292, Japan\\
$^4$Department of System Innovation, Osaka University, Toyonaka, Osaka 560-8531, Japan\\
$^5$Institute for Solid State Physics, University of Tokyo, Kashiwa 277-8581, Japan
}

\date{\today}

\begin{abstract}
A GaN film doped with 8.2 \% Mn was grown by the molecular-beam-epitaxy technique. Magnetization measurements show that this highly Mn-doped GaN film exhibits ferromagnetism above room temperature. It is also revealed that the high-temperature ferromagnetic state is significantly suppressed below 10 K, accompanied by an increase of the electrical resistivity with decreasing temperature. This observation clearly demonstrates a close relation between the ferromagnetism with extremely high-$T_{\rm C}$ and the carrier transport in the Mn-doped GaN film. 
\end{abstract}


\maketitle
%
%
The discovery of ferromagnetism in Mn-doped InAs and GaAs \cite{Munekata1, Ohno1, Ohno2} has stimulated intensive research on the III-V based diluted magnetic semiconductors (DMSs), because DMSs are considered to be potential materials for future spintronics devices which combine both charge and spin degrees of freedom in semiconductor electronics. Since practical spintronics devices are expected to work around and above room temperature, DMSs should have Curie temperatures $T_{\rm C}$ in excess of 400 K and considerable efforts have been devoted in the recent decade to synthesize such films. As to GaN-based DMSs, progress in GaN growth \cite{Strite1} and device development \cite{Mohammad1} have been remarkable in the past decade and high-conductivity, p-type epitaxial layers can be produced \cite{Amano1}. Blue-light-emitting diodes are already commercial products \cite{Nakamura1}, and blue-laser diodes \cite{Nakamura2}, ultraviolet detectors \cite{Razeghi1}, and high-power, high-temperature field-effect transistors \cite{Khan1} have been achieved and have been improved rapidly. Accordingly, the technological importance of room-temperature ferromagnetic GaN for spintronics is very large. Recently, some of the present authors have reported the successful growth of Mn-doped GaN (GaMnN) films showing room-temperature ferromagnetism \cite{Sonoda1, Sonoda2, Sonoda3, Hori1}. So far, the highest $T_{\rm C}$ is around 940 K in 5.7 \% Mn-doped GaN. Although the origin of high-temperature ferromagnetism in these GaMnN systems is still under discussion, conducting carriers are thought to play a substantial role in mediating a strong ferromagnetic interaction between the magnetic moments.

So, similar to the importance of the control of the carrier density for achieving the best semiconducting properties, the carrier density plays a crucial role in determining the magnetic properties of the material as well. Previously, we have studied the relation between the transport properties of GaMnN film and the high-$T_{\rm C}$ ferromagnetism \cite{Hori1, Miura1}. For a GaMnN film with 5.7 \% Mn, a steep decrease of the spontaneous magnetization with increasing temperature was found below 10 K, while the ferromagnetism persists up to above 750 K. Furthermore, a steep increase of the conductivity or carrier density with increasing temperature was found in the same low-temperature region. These results have been discussed in terms of the double-exchange mechanism due to the hopping electrons of the Mn-impurity band, where the decrease of the spontaneous moment with increasing temperature below 10 K was attributed to delocalization of localized carriers that also play a principal role in high-$T_{\rm C}$ ferromagnetism. These studies have well established the close relation between the carrier transport and the magnetism in high-$T_{\rm C}$ GaMnN film. However, the precise role of carrier transport in the development of the high-temperature ferromagnetic state is still unclear. 

In this paper, we clearly demonstrate evidence for carrier-induced ferromagnetism in GaN film doped with 8.2 \% Mn, grown by the molecular-beam-epitaxy (MBE) technique. Magnetization measurements point out that the GaMnN film is ferromagnetic above room temperature. Surprisingly, we found that the high-temperature ferromagnetism is significantly suppressed below 10 K, accompanied by a steep increase of the electrical resistivity. This clearly indicates the presence of weakly localized carriers in this temperature region and strongly suggests that the high-$T_{\rm C}$ ferromagnetic state is driven by carrier conduction.

%
%

The fabrication of the GaMnN film was carried out by an ammonia-MBE technique. Details of the sample preparation have been described elsewhere \cite{Sonoda1}. The wurtzite crystal structure of the obtained GaMnN film was confirmed by X-ray diffraction (XRD). The XRD study also showed that there is no secondary phase within the resolution limit. An X-ray absorption fine structure (XAFS) study confirmed that the Mn atoms are substituted for Ga atoms. By means of electron probe micro analysis (EPMA), the Mn concentration in the film was established to be 8.2 \%. 

 The temperature and field dependence of the magnetization were measured at temperatures between 2 K and 350 K by using a superconducting quantum interference device (SQUID) magnetometer (MPMS-XL, Quantum Design Co. Ltd.). The magnetic field was applied parallel to the film plane. The temperature variation of the in-plane electrical resistivity was measured by a conventional four-probe dc method.
 
 %
%

\begin{figure}[htbp]
\includegraphics[width=8cm]{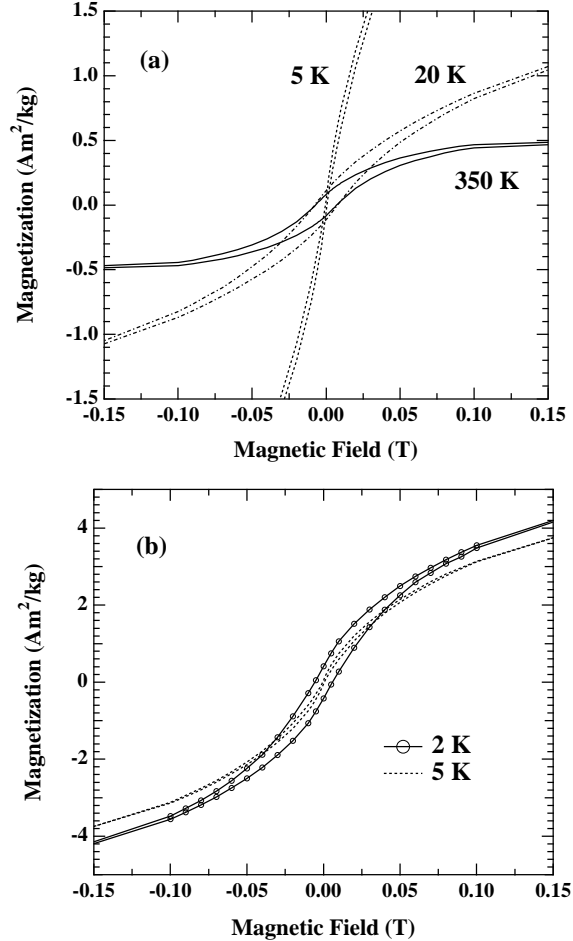}
\caption{\label{fig1} Magnetization process at various temperatures.}
\end{figure}

The field dependence of the magnetization has been investigated at various temperatures up to 350 K. Figure 1 shows representative results of $M$-$H$ curves. At fields below 0.1 T, hysteretic behavior is quite pronounced at 20 K and 350 K (Fig. 1(a)), whereas both the coercive field $B_c$ and the remanent magnetization $M_r$ are about the same at these two temperatures. This result clearly shows the high-$T_{\rm C}$ ferromagnetism of the present heavily-doped GaMnN film with a Curie temperature far above room temperature. Upon further cooling, however, the hysteresis is considerably reduced and also $M_r$ decreases as can be seen in the 5 K curve. Figure 1(b) shows that the hysteresis loop recovers again below 5 K and that $M_r$ at 2 K is several times larger than at temperatures above 20 K. Besides the ferromagnetic features evidenced by the hysteresis loop and the remanent magnetization, we see the existence of a field-dependent contribution. A similar coexistence of a field-dependent component has commonly been found in other high-$T_{\rm C}$ GaMnN films \cite{Sonoda2, Sasaki1}. Details of the magnetization behavior of the present GaMnN film in high magnetic fields will be discussed in a separate paper. Here, we will focus our attention on the low-field ferromagnetic behavior.

\begin{figure}[htb]
\includegraphics[width=8.5cm]{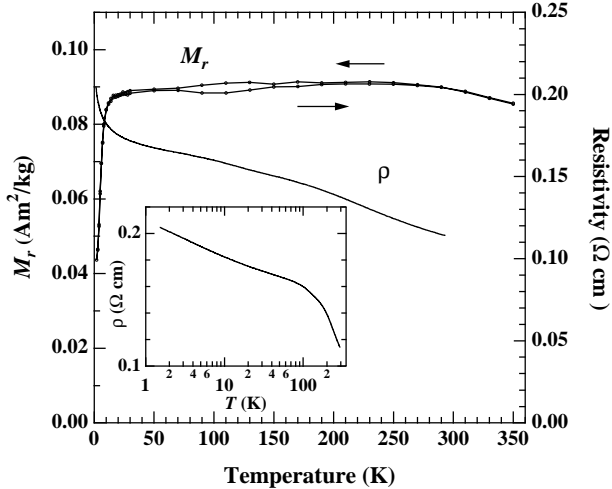}
\caption{\label{fig2} Temperature dependence of the remanent magnetization $M_r$ and the electrical resistivity $\rho$. $M_r$ and 
$\rho$ were measured after keeping the sample in applied field of 1 T for a few minutes at 350 K and 270 K, respectively. (Inset) Electrical resistivity plotted on a logarithmic temperature scale.}
\end{figure}

We further investigated the temperature evolution of the ferromagnetism of the GaMnN film by measuring the temperature dependence of the remanent magnetization $M_r$. In Fig. 2, the $M_r$-$T$ curve, as recorded during a temperature cycle 350 K $\rightarrow$ 2 K $\rightarrow$ 350 K is shown. The measurement was performed in zero field, after a field of 1 T had been applied for a few minutes at 350 K. The nearly $T$-independent behavior of $M_r$ confirms the presence of a ferromagnetic state with a $T_{\rm C}$ value of more than 350 K. Surprisingly, below about 10 K, a steep decrease of $M_r$ occurs and there is no substantial difference between the curves recorded with descending and ascending temperature. It should be noted that the difference between the two curves around 100 K is due to experimental error and is not intrinsic behavior. The reversible behavior of $M_r$ strongly suggests that the decrease of $M_r$ is not due to a domain effect. We also like to emphasize that the $T$-dependence of $M_r$ is exactly the same, at least between 5 K and 350 K, with that of the remanent magnetization $M_r$ obtained from the $M$-$H$ curves. The observed significant suppression of the high-temperature ferromagnetic state below 10 K convincingly excludes a contribution of a possible secondary phase, such as ferromagnetic Mn$_x$N$_y$ or Ga$_x$Mn$_y$ as a plausible origin of the observed high-$T_{\rm C}$ ferromagnetism, because these ferromagnets do not show such a reduction of the ferromagnetism at low temperatures \cite{Bither1, Mekata1}.

The ferromagnetism observed at high temperatures in the present GaMnN film is clearly related to the carrier transport. Figure 2 shows that the temperature dependence of the electrical resistivity $\rho$ is only gradual at temperatures above 20 K, while it shows a steep increase with decreasing temperature below 20 K. In addition, as shown in the inset of Fig. 2, below 10 K a weakly-localized nature of the carriers is suggested by the $\rho$ vs log$T$ dependence. It is worth mentioning that in previous studies \cite{Hori1, Miura1} a similar increase of $\rho$ at low temperatures was found to be accompanied with a steep decrease of the carrier density as derived from Hall measurements. Comparison of the $\rho$ vs $T$ and the $M_r$ vs $T$ curves in Fig. 2 clearly demonstrates the close relation between the transport and magnetic properties. The suppression of the high-$T_{\rm C}$ ferromagnetic state below 10 K coincides with a decrease of the carrier density which should be attributed to an increase of the localization of the carriers conduction, while the high-$T_{\rm C}$ ferromagnetic state appears in the temperature regime of delocalized carriers. We emphasize again the simultaneous decrease of $M_r$ and increase of $\rho$ at low temperatures, and the $T$-reversible behavior of these both quantities, resulting in conclusive evidence that the observed high-$T_{\rm C}$ ferromagnetism is an intrinsic property of the present GaMnN film and that the carriers play a dominant role in the high-$T_{\rm C}$ ferromagnetism.

In Figure 3, an $M_r$-$T$ curve is shown that has been obtained in a different manner. In this case, the measurement was started after applying a field of 1 T for a few minutes at 2 K and a strongly irreversible $T$ dependence of $M_r$ appears below 5 K. As shown in the inset of Fig. 3, the difference between the measurements with ascending and descending temperature, $\Delta M_r$ = $M_r$(2 K $\rightarrow$ 7 K) - $M_r$(7 K $\rightarrow$ 2 K), becomes non-zero and shows a marked increase with decreasing $T$ below 5 K. This pronounced increase of $\Delta M_r$ corresponds to the recovery of the large hysteresis loop below 5 K compared with that at high temperatures (Fig. 1). The features of $\Delta M_r$ indicate that another ferromagnetic state exists with a low $T_{\rm C}$ value. 

This existence of low-$T_{\rm C}$ ferromagnetism has also been reported for GaMnN films with other Mn concentrations of 6.8 \% \cite{Miura1} and 3 \% \cite{Sasaki1}, in which the high-$T_{\rm C}$ ferromagnetic state also persists up to above room temperature. In these studies, however, the successive suppression and enhancement of the ferromagnetic state with decreasing temperature was not reported. The reduction of the high-$T_{\rm C}$ ferromagnetism has been found for the first time in the present GaMnN film.

\begin{figure}[htb]
\includegraphics[width=8cm]{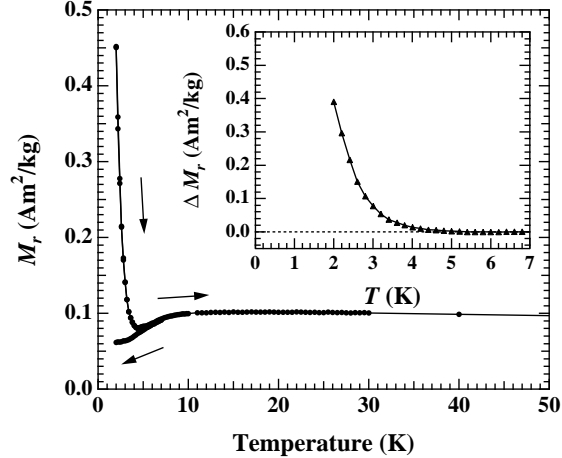}
\caption{\label{fig3} $M_r$ measured after keeping the sample in applied field of 1 T for a few minutes at 2 K for the temperature cycle 2 K $\rightarrow$ 350 K $\rightarrow$ 2 K. (Inset) The difference of $M_r$ between the ascending and descending process, $\Delta M_r$ = $M_r$(2 K $\rightarrow$ 7 K) - $M_r$(7 K $\rightarrow$ 2 K).}
\end{figure}
 
%
%

Theoretical investigations of the mechanism of the high-$T_{\rm C}$ ferromagnetism in GaMnN films have been carried out by several approaches, but the definitive answer has not yet been given. Dietl et al. \cite{Dietl1, Dietl2} have proposed a mean-field Zener p-d exchange-interaction model, which explains many characteristics of the GaMnAs system. This model is based on the idea that Mn 3d state is located within or near the valence p-band and that the polarization of the valence p-band, due to hybridization between Mn 3d-levels and p-band states, mediates the magnetic interaction between localized Mn spins. On the other hand, first-principles electronic-structure calculations have indicated that the Mn 3d state forms a pronounced impurity band in the wide band gap in GaN and that the Fermi energy is located in this impurity band \cite{Sato1, Sato2, Sanyal1, Kulatov1}. A double-exchange type of mechanism was concluded to be dominant in ferromagnetic GaMnN, with hopping Mn d-electrons in the impurity band playing a crucial role in the ferromagnetic coupling between the Mn spins. Theoretical calculations have also pointed out that the exchange interaction in GaMnN is short-range, whereas there are long-range correlations in GaMnAs.
 
 At present, the lack of detailed knowledge of the electronic structure of the studied GaMnN film prevents a substantial discussion on the interaction mechanism that leads to the ferromagnetism with high $T_{\rm C}$. However, formation of a Mn 3d impurity band in the gap was suggested by a XANES (X-ray absorption near edge structure) study, which has been performed on a GaMnN film with 5.7 \% Mn and $T_{\rm C}$ of 940 K \cite{Sonoda3}. Furthermore, a recent NEXAFS (near edge X-ray absorption fine structure) study has revealed the coexistence of Mn atoms with valencies Mn$^{2+}$and Mn$^{3+}$ in high-$T_{\rm C}$ GaMnN film \cite{Sonoda4}. With these pieces of experimental evidence, the intimate correlation between carrier trapping and the suppression of the high-$T_{\rm C}$ ferromagnetism below 10 K observed in the present study strongly suggests that the ferromagnetic coupling between Mn spins is mediated by Mn 3d electrons hopping between Mn$^{2+}$ and Mn$^{3+}$. In this scenario, the temperature evolution of the ferromagnetic state observed at low temperatures can qualitatively be understood as follows: the weak localization of hopping carriers evidenced by the $\rho$ vs log$T$ dependence below 10 K, which must be caused by a random distribution of impurity ions, considerably reduces the carrier density. Because of the short-range correlations as predicted by the theoretical calculations, reduction of the carrier density weakens the ferromagnetic interaction of the double-exchange mechanism between Mn spins, thus leading to the suppression of the high-$T_{\rm C}$ ferromagnetic state in the same temperature region. As to the low-$T_{\rm C}$ ferromagnetism, we don't have a sound explanation at present. One possible scenario is that magnetic clusters with short distance may be formed. In the absence of an applied magnetic field, the magnetic moments in the domains formed by these clusters are randomly oriented, while they are directed in the direction of the applied-field at the lowest temperature, even after the field has been turned off. In order to check this, it will be necessary to quantitatively analyze the temperature and field dependencies of magnetization, which will be presented in a future paper.
 
%
%

In summary, we have investigated the ferromagnetic properties of GaMnN film with a high $T_{\rm C}$ value. By measuring the remanent magnetization $M_r$ in different temperature cycle, it has been established that the high-$T_{\rm C}$ ferromagnetic state is considerably weakened below 10 K and that another low-$T_{\rm C}$ ferromagnetic state appears below 5 K. The observation that the localization of conducting carriers coincides with the suppression of the high-$T_{\rm C}$ ferromagnetism provides very strong evidence that the ferromagnetic interaction responsible for the high $T_{\rm C}$ value is mediated by hopping-like carrier conduction.


\emph{Acknowledgments---}
We thank Prof. F.R. de Boer for critical reading of the manuscript. This study was supported in part by a Grant-in-Aid for Scientific Research of the Ministry of Education, Culture, Sports, Science and Technology.


\end{document}